\documentclass{emulateapj}

\def\ts     {\thinspace}
\def\kms    {\ts km\ts s$^{-1}$}
\def\etal   {{\rm et\ts al.}}
\def\msol   {M$_{\odot}$}
\def\lsol   {L$_{\odot}$}

\def\aco    {$^{12}${\rm CO}($J$=1$\to$0)}

\def\dco    {$^{12}${\rm CO}($J$=4$\to$3)}

\def\cho    {{\rm H$_2$O}(3$_{13}$$\to$2$_{20}$)}

\def\bhcn    {{\rm HCN}($J$=2$\to$1)}

\slugcomment{draft version \today, accepted for publication in the Astrophysical Journal}

\shorttitle{A search for H$_2$O in MG\,0751+2716}
\shortauthors{Riechers et al.}

\begin{document}

\title{A search for H$_2$O in the strongly lensed QSO MG\,0751+2716 at 
$z$ = 3.2}

\author{Dominik A. Riechers\altaffilmark{1}, Axel Weiss\altaffilmark{2}, 
Fabian Walter\altaffilmark{1}, Christopher L. Carilli\altaffilmark{3}, 
and Kirsten K. Knudsen\altaffilmark{1}}

\altaffiltext{1}{Max-Planck-Institut f\"ur Astronomie, K\"onigstuhl
  17, Heidelberg, D-69117, Germany}

\altaffiltext{2}{Max-Planck-Institut f\"ur Radioastronomie, Auf dem
  H\"ugel 69, Bonn, D-53121, Germany}

\altaffiltext{3}{National Radio Astronomy Observatory, PO Box O,
  Socorro, NM 87801, USA}

\email{riechers@mpia.de}

\begin{abstract}
  We present a search for 183\,GHz \cho\ emission in the
  infrared--luminous quasar MG\,0751+2716 with the NRAO Very Large
  Array (VLA). At $z = 3.200 \pm 0.001$, this water emission feature
  is redshifted to 43.6\,GHz. As opposed to the faint rotational
  transitions of HCN (the standard high--density tracer at high--$z$),
  \cho\ is observed with high maser amplification factors in Galactic
  star--forming regions. It therefore holds the potential to trace
  high--density star--forming regions in the distant universe. If
  indeed all star--forming regions in massively star--forming galaxies
  at $z > 3$ have similar physical properties as e.g.\ the Orion or
  W49N molecular cloud cores, the flux ratio between the
  maser--amplified \cho\ and the thermally excited \aco\ transitions
  may be as high as factor of 20 (but has to be corrected by their
  relative filling factor).  MG\,0751+2716 is a strong \dco\ emitter,
  and therefore one of the most suitable targets to search for \cho\
  at cosmological redshifts. Our search resulted in an upper limit in
  line luminosity of $L'_{\rm H_2O} < 0.6 \times
  10^9\,$K\,\kms\,pc$^2$. Assuming a brightness temperature of $T_{\rm
    b}{\rm (H_2O)} \simeq 500$\,K for the maser emission and CO
  properties from the literature, this translates to a \cho/\dco\ area
  filling factor of less than 1\%. However, this limit is not valid if
  the \cho\ maser emission is quenched, i.e.\ if the line is only
  thermally excited. We conclude that, if our results were to hold for
  other high--$z$ sources, H$_2$O does not appear to be a more
  luminous alternative to HCN to detect high--density gas in
  star--forming environments at high redshift.
\end{abstract}

\keywords{galaxies: active, starburst, formation, high redshift --- cosmology: observations 
--- masers}

\section{Introduction}

Over the past years, massive amounts of dust and gas have been
detected in distant quasars, allowing
to study the properties of molecular gas in the early epoch of galaxy
formation.
Molecular gas in high--redshift galaxies is commonly traced by CO
emission and was found in $>$30 galaxies at $z>2$ to date, out to the
highest redshift quasar,
SDSS\,J1148+5251 at $z = 6.42$ (Walter \etal\ \citeyear{wal03},
\citeyear{wal04}; Bertoldi \etal\ \citeyear{ber03}).
The observed molecular gas masses in excess of 10$^{10}$\,\msol\
provide the requisite material for star formation (SF, e.g.\ Solomon
\& Vanden Bout \citeyear{sv05}).

The presence of abundant molecular gas, the fuel for SF, has led to
the hypothesis that the tremendous far--infrared (FIR) luminosities
($>10^{12}$\,\lsol) of these high--redshift objects are not only
powered by active galactic nuclei (AGN) but also by major starbursts
(SBs), suggesting that this population represents the formation of
large spheroidal galaxies beyond redshift 2 (Blain \etal\
\citeyear{bla02}).
However, the relative contribution to the FIR luminosity from dust
heated by SF and AGN activity in high--$z$ sources remains subject to
discussion (e.g.\ Andreani \etal\ \citeyear{and03}).

While the lower--$J$ transitions of CO are a good indicator for the
total molecular gas content of a system (e.g.\ Carilli \etal\
\citeyear{car99}, \citeyear{car02}; Riechers \etal\ \citeyear{rie05}),
they can be excited at relatively low densities; the critical density
lies at only $n_{\rm H_2} \simeq 10^3\,$cm$^{-3}$. Hence, it is a
relatively poor tracer of the denser gas directly involved in massive
star formation. The standard tracer of the dense molecular gas phase
is HCN, and the critical density to excite its lower--$J$ transitions
is $n_{\rm H_2} \simeq 10^5$ cm$^{-3}$ (due to a higher dipole moment
in comparison to CO). In this context, recent studies of the dense
molecular gas phase in local ($z<0.3$) luminous and ultra--luminous
infrared galaxies (LIRGs/ULIRGs) have shown a correlation between the
HCN luminosity and the star--formation rate (SFR) as traced by the FIR
luminosity. This correlation is much tighter than that between the CO
and FIR luminosities (Solomon \etal\ \citeyear{sol92a}; Gao \& Solomon
\citeyear{gao04a}, \citeyear{gao04b}).
Thus, observations of HCN towards high--$z$ sources hold the potential
to pin down the contribution of SF to the total FIR luminosity.
Unfortunately, emission lines from HCN in LIRGs/ULIRGs are typically
by a factor of 4 to 10 fainter than those of CO (even a factor of
25--40 in ordinary spiral galaxies, Gao \& Solomon \citeyear{gao04b}),
which hinders systematic HCN surveys at cosmological distances using
current telescopes.  Until today, deep observations of HCN at $z>2$
resulted in four detections and four upper limits (Solomon \etal\
\citeyear{sol03}; Vanden Bout \etal\ \citeyear{vdb04}; Carilli \etal\
\citeyear{car04}; Wagg \etal\ \citeyear{wag05}).

Due to the faintness of emission connected with thermally excited
rotational transitions of HCN, it is important to investigate whether
another physical process can be found which produces emission lines
with significantly higher luminosities.
The 183\,GHz $3_{13} \to 2_{20}$ emission line of para--H$_2$O
holds the potential to be such a tracer: in warm, dense SF regions
($n_{\rm H_2} > 10^5$\,cm$^{-3}$), this line is collisionally pumped
at relatively low kinetic temperatures ($T_{\rm kin} = 50 - 100\,$K,
e.g. Cernicharo \etal\ \citeyear{cer94}).  In contrast to compact
H$_2$O maser sources at 22\,GHz, the 183\,GHz maser emission is
spatially extended in high--mass (Orion, Cernicharo \etal\
\citeyear{cer94}; W49N, Gonz{\'a}lez-Alfonso \etal\ \citeyear{gon95};
Sgr\,B2, Cernicharo \etal\ \citeyear{cer06}) and even low--mass
(HH7--11, Cernicharo \etal\ \citeyear{cer96}) star--forming regions.
Due to maser amplification, the observed 183\,GHz \cho\ line
brightness temperatures in these regions exceed those observed in
\aco\ by up to a factor of 20 (e.g.\ for Orion, Cernicharo \etal\
\citeyear{cer94}; Schulz \etal\ \citeyear{sch95}).

However, the \cho\ line is strongly absorbed by the terrestrial
atmosphere, rendering detection in nearby galaxies very difficult. The
first successful extragalactic detection of \cho\ has been reported in
NGC\,3079 (Humphreys \etal\ \citeyear{hum05}). Previous searches in
low--$z$ (U)LIRGs have only resulted in upper limits (Combes \etal\
\citeyear{com97}).  Recently, the \cho\ transition has also been
detected towards Arp\,220 (Cernicharo, Pardo \& Weiss 2006, in prep.,
see Table 1).
At high redshift, only 
the $2_{11} \to 2_{02}$ transition of para--H$_2$O at 752.033\,GHz
(rest frame) has tentatively been detected in the $z \simeq 2.3$ QSO
IRAS\,F10214+4724 (Encrenaz \etal\ \citeyear{enc93}; Casoli \etal\
\citeyear{cas94}).

\indent In this paper, we report on a search for maser emission from
the 183\,GHz water line in the strongly lensed, radio--loud $z = 3.2$
QSO MG\,0751+2716 using the VLA\footnote{The Very Large Array is a
  facility of the National Radio Astronomy Observatory, operated by
  Associated Universities, Inc., under a cooperative agreement with
  the National Science Foundation.}.
Due to its strong magnification, MG\,0751+2716 is the brightest CO
source at high redshift that can be observed in the \cho\ transition
with the VLA. Its $L'_{\rm CO}$ and $L_{\rm FIR}$ (see Table 1) are
comparable to local starburst galaxies like Arp\,220. This suggests
that this source is undergoing massive star--formation, which
contributes significantly to its far--IR luminosity (e.g.\ Fig.\ 8 in
Solomon \& Vanden Bout \citeyear{sv05}).
We use a standard concordance cosmology throughout, with $H_0 =
71\,$\kms\,Mpc$^{-1}$, $\Omega_{\rm m} =0.27$, and $\Omega_{\Lambda} =
0.73$ (Spergel \etal\ \citeyear{spe03}).

\section{Observations}

We observed the \cho\ transition ($\nu_{\rm rest} = 183.3101\,$GHz)
towards MG\,0751+2716 using the VLA in D configuration on 2004 June 19
and 27. At the target $z$ of 3.200, this transition is redshifted to
43.6453\,GHz (6.87\,mm). The total on--sky integration time amounts to
16\,hr. Observations were performed in fast--switching mode using the
nearby source 0745+241 for secondary amplitude and phase calibration.
Observations were carried out under good weather conditions with 26
antennas.  The phase stability in all runs was good (typically
$<$25$^\circ$ rms for the longest baselines). For primary flux
calibration, 3C286 was observed during each run.

Two 50\,MHz (corresponding to 344\kms\ at 43.6\,GHz) intermediate
frequencies (IFs) were observed simultaneously in the so--called
'quasi--continuum' mode.
One IF was centered on the line frequency at 43.6351\,GHz (closest
possible tuning frequency, offset by $\sim$70\kms\ from the \dco\ line
center of Barvainis \etal\ \citeyear{bar02}), and the second IF was
centered at two frequencies symmetrically offset by 150\,MHz
(1030\kms, i.e.\ at 43.4851\,GHz and 43.7851\,GHz) from the line
frequency to monitor the 7\,mm continuum of MG\,0751+2716
simultaneously. Despite the small offset from the line center, the
frequency/velocity coverage is well--matched to a line with a FWHM
similar to the \dco\ line reported by Barvainis \etal\
(\citeyear{bar02}).

For data reduction and analysis, the
${\mathcal{AIPS}}$\footnote{www.aoc.nrao.edu/aips/} package was used.
The two continuum channels were concatenated in the uv/visibility
plane. The data were mapped using the CLEAN algorithm and 'natural'
weighting; this results in a synthesized beam of
1.8\,$''$$\times$1.7\,$''$ ($\sim$13\,kpc at $z = 3.2$) at a major
axis position angle of 78$^\circ$.  The final rms in both the line
channel and the combined continuum channel is 60\,$\mu$Jy beam$^{-1}$.
\section{Results}

In the top panel of Fig.\ 1, the final map of the 'ON' channel
(line+continuum emission) is shown, while the middle panel shows the
map of the two combined continuum ('OFF') channels, representing the
continuum--only flux.  The source is clearly detected in both maps. To
derive the total flux density of the source, a 2--dimensional Gaussian
was fitted to the detected source structure. Our best fit for
MG\,0751+2716 gives a Gaussian diameter of 2.3\,$''$$\times$2.3\,$''$
(source convolved with the synthesized beam); i.e.\ the emission
appears only marginally resolved in our observations, as our beam is
too large to recover the lens image substructure seen at higher
resolution (Carilli \etal\ \citeyear{car04}).  The integrated flux
density in the 'ON' channel is 10.3 $\pm$ 0.2\,mJy (peak: 6.12 $\pm$
0.06\,mJy beam$^{-1}$), and the continuum flux in the combined 'OFF'
channels is 10.3 $\pm$ 0.2\,mJy (peak: 5.88 $\pm$ 0.06\,mJy
beam$^{-1}$). From the VLA 42.2\,GHz measurement of 13.2\,mJy (Carilli
\etal\ \citeyear{car04}) and using a spectral index of $\beta = -1.2$
(Leh{\'a}r \etal\ \citeyear{leh97}), an extrapolated flux density of
12.7\,mJy can be calculated, which is higher than derived from our
observations. Therefore, continuum variability cannot be excluded.

In the bottom panel of Fig.\ 1, the continuum-subtracted 'line' map is
shown. This map was generated by subtracting a CLEAN component model
of the continuum emission (combined 'OFF' channel) from the visibility
data of the 'ON' channel, which was then imaged applying the CLEAN
algorithm with the same parameters used to create the continuum model.
No clear evidence for \cho\ line emission is found.  Thus, we set a
3$\sigma$ upper limit of $S_{\rm H_2O} < 180\,\mu$Jy on the peak flux
of \cho\ emission.  For MG\,0751+2716, we thus derive an upper limit
for the \cho\ line luminosity of $L'_{\rm H_2O} < 0.6 \times
10^9\,$K\,\kms\,pc$^2$ (3$\sigma$ limit, corrected for gravitational
magnification, $\mu_{\rm L} = 17$, Barvainis \etal\ \citeyear{bar02},
cf.\ Table \ref{tab-1}).

\section{Discussion}

\subsection{Intrinsic line brightness temperatures} \label{temp}

Given that the 183\,GHz water maser line most likely arises from
extended regions associated with star formation (Cernicharo \etal\
\citeyear{cer94}), we can use our upper limit together with the \dco\
line luminosity (Barvainis \etal\ \citeyear{bar02}) to estimate how
much of the molecular gas in MG\,0751+2716 may be associated with
massive star forming regions. Radiative transfer models based on the
\cho\ emission line show that it is inverted in warm and dense
environments ($T_{\rm kin}>40$\,K, $n_{\rm H_2} > 5 \times
10^{4}$\,cm$^{-3}$) which are typically found in star--forming
regions, and that the resulting line brightness temperature critically
depends on the underlying physical conditions.  It is known to range
from thermalized emission up to line temperatures of $\sim$10000\,K
(Cernicharo \etal\ \citeyear{cer94}; Combes \etal\ \citeyear{com97}).
As an example, observations of the 183\,GHz water maser towards Orion
IRc2 show that emission arises from narrow (few \kms) line features
with up to $\simeq 2000$\,K as well as a broad ($\simeq 200$\,\kms)
emission plateau with $\simeq 500$\,K peak line brightness temperature
(Cernicharo \etal\ \citeyear{cer94}, \citeyear{cer99}). As an order of
magnitude estimate, we assume in the following that the \cho\ line in
MG\,0751+2716 arises from a similar environment as Orion with an
intrinsic brightness temperature of $T_{\rm b}{\rm (H_2O)} \simeq
500$\,K.  Only the \dco\ line is observed so far in MG\,0751+2716,
therefore the brightness temperature of the CO emission is not
well--constrained. However, the dust continuum temperature of $T_{\rm
  dust}\simeq 40$\,K in MG\,0751+2716 (Barvainis \& Ivison
\citeyear{bi02}) is consistent with those observed in most other
high--$z$ QSOs (e.g.\ Beelen et al.\ \citeyear{bee06}).  In addition,
the CO brightness temperature in well--studied high--$z$ objects such
as IRAS\,F10214+4724 agrees within a factor of $\sim 2$ with the dust
temperature (e.g. Downes \etal\ \citeyear{dow95}). By analogy, we here
assume $T_{\rm b}{\rm (CO)}\simeq 40$\,K.

\subsection{Area filling factor} \label{filf}

The combined ratios of the brightness temperatures and line
luminosities of H$_2$O and CO ($T_{\rm b}{\rm (H_2O)} \simeq 500$\,K,
$T_{\rm b}{\rm (CO)}\simeq 40$\,K, $L'_{\rm H_2O} < 0.6 \times
10^9\,$K\,\kms\,pc$^2$, $L'_{\rm CO} = 10.0 \times
10^9\,$K\,\kms\,pc$^2$) can be used to estimate the relative area
filling factor (FF) of both molecules.  By this means, our observed
upper limit on the 183\,GHz H$_2$O line luminosity together with the
\dco\ line luminosity translates into an upper limit on the H$_2$O/CO
area filling factor of $<1\%$.

To better understand the relevance of this filling factor limit
derived from our observations, we now give an independent estimate for
the filling factor based on the FIR luminosity and geometrical
arguments. For this purpose, we assume in the following that all
183\,GHz H$_2$O emission in MG\,0751+2716 is created in hot,
star--forming cores like the one found in the central region of Orion.
Assuming $L_{\rm FIR} = 1.2 \times 10^{5}$\,\lsol\ for the central
1\,$\square '$ of the Orion Nebula (Werner \etal\ \citeyear{wer76},
corresponding to an area of 0.013\,pc$^2$ at a distance of 450\,pc),
we find that 10$^7$ of these cores are needed to account for the
far--IR luminosity in MG\,0751+2716 (Table 1). Assuming a size of
0.013\,pc$^2$ each, 10$^7$ of such cores would fill a disk with an
equivalent radius of 200\,pc. Taking the observed \dco\ line FWHM and
luminosity into account, and assuming $T_{\rm b}{\rm (CO)}\simeq
40$\,K, for MG\,0751+2716, we can calculate that the {\em dust and CO}
are distributed over a region with an equivalent radius of 450\,pc.
Comparing the derived disk sizes, we thus find an expected H$_2$O/CO
area filling factor of $\sim$20\%.  This estimate is much higher than
what is actually observed.

\subsection{Comparison with other (U)LIRGs and NGC\,3079} \label{comp}

Our results for MG\,0751+2716 are in line with observations of the
radio--quiet QSOs Mkn\,1014 (PG\,0157+001, $z=0.16$) and VII\,Zw\,244
(PG\,0838+770, $z=0.13$, see Table 1), the only two other
infrared--luminous galaxies for which upper limits on the 183\,GHz
H$_2$O line have been reported (Combes \etal\ \citeyear{com97}) to
date. Using the same assumptions on the intrinsic line brightness
temperatures as for MG\,0751+2716, the non--detections in these
galaxies imply filling factors of the water line relative to \aco\ of
$<5\%$ and $<2\%$ for Mkn\,1014 and VII\,Zw\,244 respectively. As CO
emission appears to be close to thermalized up to the \dco\ transition
in high--$z$ QSOs (Weiss et al.\ \citeyear{wei05}), these limits are
directly comparable to MG\,0751+2716.

In Arp\,220 ($z=0.018$), the \cho\ line was detected with a line
luminosity of $L'_{\rm H_2O} = 2.9 \times 10^8\,$K\,\kms\,pc$^2$
(Cernicharo, Pardo \& Weiss 2006, in prep.). With the CO luminosity
given in Solomon \etal\ (\citeyear{sol92a}), this translates to a
H$_2$O/CO filling factor of 0.4\%. Arp\,220 has a far--IR luminosity
similar to that of MG\,0751+2716, which might suggest a similar
H$_2$O/CO filling factor.
This suggests that the \cho\ line might be detectable in MG\,0751+2716
if the sensitivity was increased only by a factor of few.

For the recent first extragalactic \cho\ detection in the LINER
NGC\,3079 (Humphreys \etal\ \citeyear{hum05}), the H$_2$O/CO
luminosity ratio (using same brightness temperatures as before)
translates into a H$_2$O/CO area filling factor of only $0.05\%$.
However, it is difficult to assess the relevance of the NGC\,3079
results for MG\,0751+2716, as the former is not a ULIRG. For
lower--luminosity galaxies, it has been found that the HCN/CO
luminosity ratio is significantly lower (2.5--4\% rather than
10--25\%, Gao \& Solomon \citeyear{gao04b}); by analogy, also lower
H$_2$O/CO luminosity ratios may be expected in galaxies with FIR
luminosities that are significantly lower than in a ULIRG.

\subsection{Possible interpretations of the non--detection} \label{conc}

Our observations suggest that the H$_2$O/CO area filling factor in
MG\,0751+2716 is significantly lower than expected from simple
estimates based on the distribution and temperature of the dust and
CO.

However, other reasons for the weakness of the water emission should
also be considered: as discussed in Combes \etal\ (\citeyear{com97}),
the physical conditions to efficiently pump the 183\,GHz line to the
assumed 500\,K may not be present.  There are strong differences in
the conditions for H$_2$O line emission among Galactic star--forming
regions.  E.g., the \cho\ maser emission in Sgr\,B2(N/M) is much
weaker than that in Orion\,IRc2.  The extent of the maser emitting
regions in both sources is of order 1\,pc, but the observed brightness
temperature in Sgr\,B2 is by 1--2 orders of magnitude lower, reaching
only few tens of Kelvin
(Cernicharo \etal\ \citeyear{cer94}, \citeyear{cer06}).
This is much lower than the 500\,K we assumed for a typical H$_2$O
emitting region.  The 183\,GHz H$_2$O lines are thus only by a factor
of 3 amplified relative to CO in this source.  For high densities
and/or column densities, which are known to be present in the compact
molecular distributions of luminous IR galaxies, the maser emission of
the 183\,GHz line may even be quenched. This would lead to thermalized
emission without maser amplification (Cernicharo \etal\
\citeyear{cer94}; Combes \etal\ \citeyear{com97}). For thermalized
emission in an optically thick environment, the \cho\ line is not
expected to be stronger than those of other high density gas tracers
such as HCN. In Arp\,220, the HCN luminosity is $L'_{\rm HCN} = 9.4
\times 10^8\,$K\,\kms\,pc$^2$, i.e.\ by a factor of 3 brighter than
\cho\ (Cernicharo, Pardo \& Weiss 2006, in prep.; Graci\'a-Carpio
\etal\ \citeyear{gra06}).  The H$_2$O and HCN luminosities in Arp\,220
would thus even be consistent with $T_{\rm b}{\rm (H_2O)} = T_{\rm
  b}{\rm (HCN)}$ and an area filling factor that is smaller for H$_2$O
than for HCN.  This suggests that only a small fraction of the dense
gas is giving rise to H$_2$O emission.
Carilli \etal\ (\citeyear{car04}) set a 3$\sigma$ limit of $L'_{\rm
  HCN} < 1.0 \times 10^9\,$K\,\kms\,pc$^2$ on the \bhcn\ line
luminosity in MG\,0751+2716, which is twice as high as our limit for
$L'_{\rm H_2O}$.

It also remains a possibility that the reason for our non--detection
is due to limitations in our observing mode.  Due to the small
fraction of gas giving rise to the \cho\ emission compared to CO, its
linewidth could be much smaller than that of the \dco\ transition.
This has recently been found in NGC\,3079 (Humphreys \etal\
\citeyear{hum05}), where the difference in linewidth is more than a
factor of 10.  As our observations had to be done using 50\,MHz
(344\,\kms) channels, the emission from a narrow line would be diluted
over the full velocity range covered by that channel, rendering
detection unlikely even at the achieved sensitivity. However, in
Arp\,220, which is likely more similar to MG\,0751+2716, the \cho\
line has a width of 350\,\kms. Such a linewidth would match our
observing mode very well.
Finally, differential lensing may have an impact on the measured
luminosity ratios ($\mu_{\rm L} = 17$), in particular if the CO and
H$_2$O emission do not emerge from the same regions.

Given the aforementioned results, we find that water maser activity
does not outshine emission from rotational transitions of CO or HCN in
the $z=3.2$ QSO MG\,0751+2716.  If our results can be generalized, we
conclude that H$_2$O (despite the fact that it potentially traces hot,
star--forming cores) does not appear to be a more luminous alternative
to HCN to detect high--density gas in star--forming environments at
high redshift.

\acknowledgments 
The National Radio Astronomy Observatory is operated by Associated
Universities Inc., under cooperative agreement with the National
Science Foundation.  D.~ R.\ acknowledges support from the Deutsche
Forschungsgemeinschaft (DFG) Priority Programme 1177.  C.~C.\
acknowledges support from the Max-Planck-Gesellschaft and the
Alexander von Humboldt-Stiftung through the
Max-Planck-Forschungspreis.  The authors would like to thank the
referee for useful comments which helped to improve the manuscript.



\begin{deluxetable}{ l c c c c c c c c c }
\tabletypesize{\scriptsize}
\tablecaption{Extragalactic para--\cho\ emission: fluxes, luminosities, and area filling factors.\label{tab-1}}
\tablehead{
& $z$ & $D_{\rm L}$& $\nu_{\rm obs}$ & $S_{\rm H_2O}$ & $L'_{\rm CO}$\tablenotemark{a} & $L'_{\rm H_2O}$\tablenotemark{a} & $L_{\rm FIR}$ & FF & Ref. \\
& & [Mpc] & [GHz] & [mJy] & [10$^9$\,K\,\kms\,pc$^2$] & [10$^9$\,K\,\kms\,pc$^2$] & [10$^{12}\,$\lsol] & & }
\startdata
MG\,0751+2716\tablenotemark{b} & 3.200  & 27940 & 43.6453  & $<$0.18 & 10.0 & $<0.6$ & 1.2  & $<$1\% & 1,2,3 \\
Mkn\,1014     & 0.1631 & 774   & 157.6048 & $<$26   & 7.8 & $<2.9$ & 2.2  & $<$5\% & 4,5 \\
VII\,Zw\,244  & 0.1324 & 616   & 161.8775 & $<$26   & 3.9 & $<0.7$ & 0.14 & $<$2\% & 4,5 \\ 
\tableline
NGC\,3079     & 0.003723 & 16 & 182.6302 & 550 & 1.1 & 0.0041 & 0.021 & 0.05\% & 6,7 \\
Arp\,220      & 0.018126 & 78 & 180.0680 & 170 & 5.9 & 0.29 & 1.3 & 0.4\% & 8,9,10 \\
\vspace{-3mm}
\enddata 
\tablenotetext{a}{Derived as described by Solomon \etal\ (\citeyear{sol92b}): 
$L'_{\rm X}[{\rm K \ts km\ts s^{-1} pc^2}] = 3.25 \times 10^7 \times I \times \nu_{\rm obs}^{-2} \times D_{\rm L}^2 \times (1+z)^{-3}$, 
where X is the molecule, $I$ is the velocity--integrated line flux in Jy \kms, $D_{\rm L}$ is the luminosity
distance in Mpc, and $\nu_{\rm obs}$ is the observed frequency in GHz.}
\tablenotetext{b}{This QSO is lensed by a factor of $\mu_{\rm L} = 17$ (Barvainis \etal\ \citeyear{bar02}). All given luminosities are corrected for lensing. }
\tablerefs{${}$[1] This work, [2] Barvainis \etal\ (\citeyear{bar02}), [3] Carilli \etal\ (\citeyear{car04}), 
[4] Combes \etal\ (\citeyear{com97}), [5] Alloin \etal\ (\citeyear{all92}), 
[6] Koda \etal\ (\citeyear{kod02}), [7] Humphreys \etal\ (\citeyear{hum05}),
[8] Cernicharo, Pardo \& Weiss (2006), in prep., [9] Solomon \etal\ (\citeyear{sol92a}), [10] Solomon \etal\ (\citeyear{sol97}).}
\tablecomments{For MG\,0751+2716, the \dco\ FWHM linewidth 
(Barvainis \etal\ \citeyear{bar02}, corrected for the VLA bandpass) of 350\,\kms\ is utilized to derive $L'_{\rm H_2O}$, whereas the 
\aco\ FWHM linewidths of 210\,\kms\ and 80\,\kms\ (Combes \etal\ \citeyear{com97}) are assumed for Mkn\,1014 and VII\,Zw\,244.}
\end{deluxetable}



\begin{figure}
\epsscale{.5}
\plotone{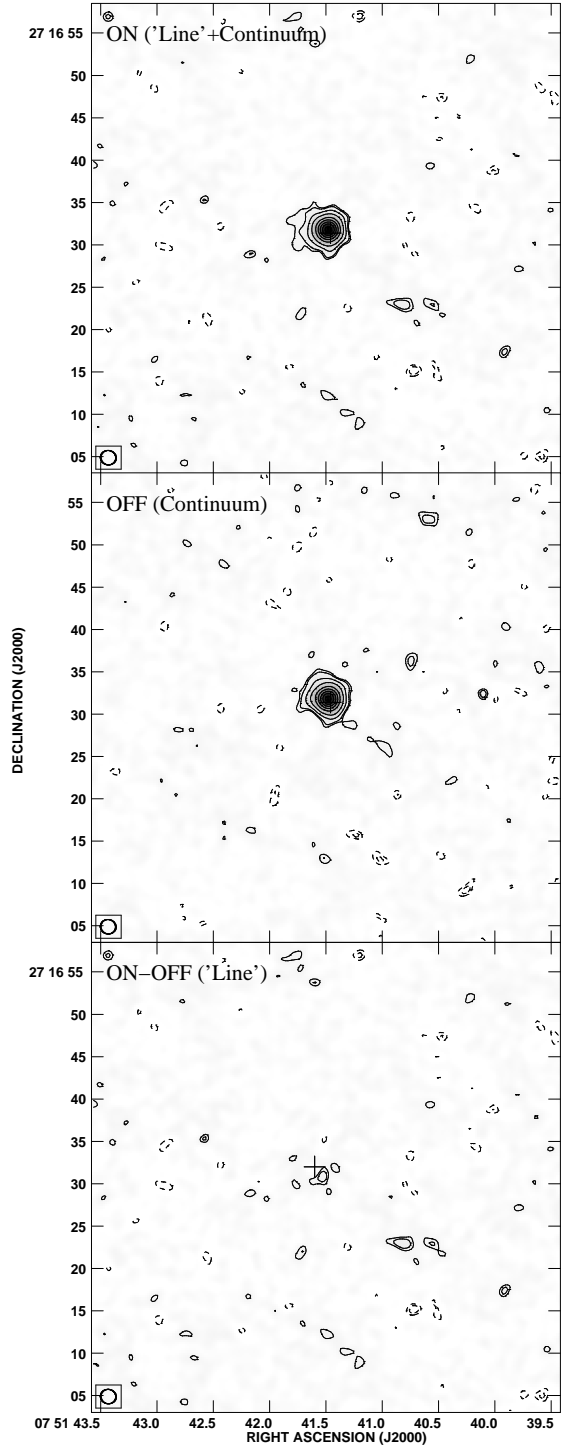}
\caption{VLA observations of MG\,0751+2716 at a resolution of 1.8\,$''$$\times$1.7\,$''$ at a major axis position angle of 
78$^\circ$. The circle in the bottom left corner of each map represents the FWHM of the restoring CLEAN beam. 
\textbf{Top}: Map of the central 50\,MHz (344\kms) 'ON' channel at 43.6\,GHz (183.3\,GHz rest--frame),
the frequency for which the H$_2$O line is expected. \textbf{Middle}: 
Combined map of the two 50\,MHz 'OFF' (continuum only) channels symmetrically offset by 150\,MHz (1030\kms) each from 
the H$_2$O line frequency. The source is marginally resolved in both maps. 
\textbf{Bottom}: Continuum--subtracted ('ON--OFF') map. No clear evidence for \cho\ line emission is found 
within the uncertainties of the observations. 
All three maps are shown with (-4, -3, 3, 4, 8, 16, 32, 48, 64, 80, 96)$\times 60\,\mu$Jy beam$^{-1}$ contours.  
 \label{f1}}
\end{figure}
\end{document}